\documentclass[aps,prl,groupedaddress,showpacs,preprint]{revtex4-1}
\usepackage{graphicx}
\usepackage{helvet}
\usepackage{amsmath,amssymb}
\usepackage{bm}
\usepackage{mathrsfs}

\begin{document}

\title{Spin-wave spectroscopy on Dzyaloshinskii-Moriya interaction\\
in room-temperature chiral magnets hosting skyrmions}
\author{R. Takagi$^{1}$, D. Morikawa$^{1}$, K. Karube$^{1}$, N. Kanazawa$^{2}$, K. Shibata$^{1}$,\\
G. Tatara$^{1}$, Y. Tokunaga$^{3}$, T. Arima$^{1,3}$, Y. Taguchi$^{1}$, Y. Tokura$^{1,2}$, and S. Seki$^{1}$}
\affiliation{$^{1}$RIKEN Center for Emergent Matter Science(CEMS), Wako 351-0198, Japan,\\
$^{2}$Department of Applied Physics, University of Tokyo, Bunkyo-ku 113-8656, Japan, $^{3}$Department of Advanced Materials Science, University of Tokyo, Kashiwa 277-8561, Japan}
\date{\today}

\begin{abstract}
Propagation character of spin wave was investigated for chiral magnets FeGe and Co-Zn-Mn alloys, which can host magnetic skyrmions near room temperature.
On the basis of the frequency shift between counter-propagating spin waves, the magnitude and sign of Dzyaloshinskii-Moriya (DM) interaction were directly evaluated.
The obtained magnetic parameters quantitatively account for the size and helicity of skyrmions as well as their materials variation, proving that the DM interaction plays a decisive role in the skyrmion formation in this class of room-temperature chiral magnets.
The propagating spin-wave spectroscopy can thus be an efficient tool to study DM interaction in bulk single-phase compounds.
Our results also demonstrate a function of spin-wave diode based on chiral crystal structures at room temperature.
\end{abstract}

\maketitle

Recently, the concept of magnetic skyrmions, i.e. vortex-like swirling spin texture with topologically stable particle nature, has attracted much attention as potential information carriers for novel magnetic information storage and processing devices \cite{Bogdanov,Rosler,1,2,Yu_nature,Fert,Sampaio,7}.
The skyrmion and associated helical spin texture can be stabilized by several distinctive mechanisms, such as Dzyaloshinskii-Moriya interaction (DMI) \cite{Bogdanov,Rosler}, frustrated exchange interactions \cite{frustration,Hayami} or the competition between magnetic dipole-dipole interaction and magnetic anisotropy \cite{Bubble,dipoleIsing}.
So far, the experimental observation of skyrmions has mainly been reported for a series of noncentrosymmetric ferromagnets, where the sizable contribution of DMI is expected \cite{1,Yu_nature,Cu2OSeO3,FeGe_LTEM,Tokunaga}.
However, the full understanding for DMI in the metallic system is often more difficult than the case for the insulating system, and recent theories suggest its relevance to quantum Berry phase and band anti-crossing that causes the complicated $E_{\rm F}$ (Fermi energy)-dependence of DMI \cite{BerryDM,Koretsune,Kikuchi}.
To unambiguously elucidate the microscopic origin of skyrmion formation for each compound, the direct quantitative evaluation of relevant magnetic parameters, in particular the magnitude and sign of DMI, is important.

To directly evaluate DMI, one promising approach is the analysis of spin wave dispersion in the ferromagnetic state.
It is generally symmetric (i.e. even function) with respect to the wave number $k$, but can become asymmetric only under the existence of $k$-linear term originating from DMI that causes the energy shift between ${\pm}k$ \cite{Kataoka}.
The direct observation of DMI based on this idea has recently been reported for the interface of bilayer films by employing several surface-sensitive methods such as Brillouin light scattering \cite{BLS1,BLS2} and spin-polarized electron energy loss spectroscopy \cite{SPEELS}.
For bulk single-phase compounds, on the other hand, the direct quantitative evaluation of DMI has rarely been reported.
Only recently, the alternative method based on neutron inelastic scattering technique \cite{TJSato} and propagating spin-wave spectroscopy (PSWS) \cite{Vlaminck1,Vlaminck2,NanoLett,Iguchi_Onose,Seki} has been proposed.

Among a series of single-phase compounds hosting magnetic skyrmions, the most promising ones for potential application are chiral-lattice helimagnetic metals FeGe ($T_c = 280$ K) \cite{FeGe_LTEM,FeGe_SANS,acchi_FeGe,FeGe_polSANS} and Co-Zn-Mn alloys ($T_c > 400$ K) \cite{Tokunaga,Karube,CoZnMn}, which are characterized by exceptionally high magnetic ordering temperature $T_c$.
In particular, Co-Zn-Mn alloys can host various unique forms of skyrmions, such as triangular and squaric lattice forms as well as highly distorted forms, for a wide compositional range \cite{Tokunaga,Karube}, and would allow the fine tuning of size and stability of skyrmions with alloy composition.
For such an essential class of materials, however, the quantitative evaluation of associated magnetic interactions is still lacking.

In this Letter, we investigated the relevant magnetic parameters for such room-temperature chiral-lattice magnets FeGe, Co$_{8}$Zn$_{8}$Mn$_{4}$ and Co$_{9}$Zn$_{9}$Mn$_{2}$ by using propagating spin-wave spectroscopy.
The spin wave in the ferromagnetic state in chiral materials shows nonreciprocal propagation, namely accompanying a nonzero frequency shift dependent on the propagation direction; this allows the direct evaluation of the magnitude and sign of DMI.
In combination with the real-space observation of spin texture, our analysis quantitatively proved that the DMI plays a decisive role in the size and helicity of skyrmions as well as their material systematics.

The crystal structures of FeGe and Co-Zn-Mn alloys are indicated in Figs. 1(g) and (h).
They possess neither inversion center nor mirror plane, and belong to chiral cubic space groups $P2_{1}3$ and $P4_{1}32$, respectively.
Their reported $H$(magnetic field)-$T$(temperature) phase diagrams are summarized in Figs. 1 (a)-(c) \cite{acchi_FeGe,Tokunaga}.
In all compounds, the helical spin order is realized under zero magnetic field, while the formation of skyrmion lattice has been reported for narrow $T$ regions just below $T_{c}$.
By applying a magnetic field larger than a critical value $H_{c}$, the collinear ferromagnetic spin state is induced.
Our measurements of spin wave propagation character were always performed in the $H$-induced ferromagnetic state.

Single crystals of FeGe and Co-Zn-Mn alloys were grown by chemical vapor transport method \cite{FeGe_CVT} and Bridgman method \cite{CoZnMn}, respectively.
The operating principle of PSWS is described in Refs. \cite{Seki,Vlaminck1,Vlaminck2}; Figure 1(i) indicates the device structure employed for this measurement.
Spin waves were emitted and detected by a pair of coplanar waveguides, which are located below a plate-shaped single crystal of FeGe or Co-Zn-Mn alloy with a typical thickness of 1 ${\mu}$m.
Injection of oscillating electric current $I^{\nu}$ into one waveguide generates an oscillating magnetic field $H^{\nu}$ and excites spin wave modes, which propagate through the crystal and induce additional electric voltage on each waveguide.
By analyzing the magnetic resonance behavior in the reflectance ${\Delta} S_{11}$ and mutual inductance ${\Delta} L_{mn}$ (with $n$ and $m$ being port number (1 or 2) used for the excitation and detection of spin wave, respectively), we can evaluate the local excitation strength and propagation character of spin wave, respectively.
Here, the wave vector ${\vec k}$ of spin wave is along the $[110]$ axis for FeGe and along the $[100]$ axis for Co-Zn-Mn alloys, respectively, and external magnetic field is always applied parallel to it (i.e. $H||k$), corresponding to backward volume wave geometry \cite{MagnetostaticWave}.
The wave number $k$ of the excited spin wave is determined by the Fourier transform of the waveguide pattern \cite{Vlaminck1,Vlaminck2}.
Unless specified, we employed the waveguide pattern with the periodicity of ${\lambda} = 12$ ${\mu}$m and the propagation gap $d=20$ ${\mu}$m characterized by a main peak of wave number distribution at $k_{\rm p} = 2{\pi}/{\lambda} = 0.50$ ${\mu}$m$^{-1}$ (See Supplemental Material \cite{SM} for details).

First, we investigated the magnetic field dependence of microwave absorption spectra $|{\Delta}S_{11}|$ for FeGe, Co$_{8}$Zn$_{8}$Mn$_{4}$ and Co$_{9}$Zn$_{9}$Mn$_{2}$ at 200 K (Figs. 1(d)-(f)), which represents the magnitude of local spin wave excitation.
In the helical spin state below critical magnetic field  $H_{c}$, the application of magnetic field gradually suppresses the magnetic resonance frequency.
In the ferromagnetic state above $H_{c}$, by contrast, the magnetic resonance frequency linearly increases as a function of $H$.
Such behaviors are commonly found for all three compounds, in accord with the theoretical predictions \cite{Seki,Onose,Schwarze,Okamura,Kataoka}.

Next, we investigated the propagation character of spin wave in the collinear ferromagnetic state for each compound. 
First, we focus on the case for FeGe.
Figure 2(a) shows the spectra of the imaginary part of the mutual inductance Im[${\Delta}L_{21}$] and Im[${\Delta}L_{12}$] measured at 200 K under $H=+300$ mT (i.e. in the collinear ferromagnetic spin state).
Here, ${\Delta}L_{21}$ and ${\Delta}L_{12}$ represent the spin wave propagating with the wave vector $+k$ and $-k$, respectively, and the corresponding experimental configuration is illustrated in the inset.
Both Im[${\Delta}L_{21}$] and Im[${\Delta}L_{12}$] show spin wave signals around the magnetic resonance frequency, but with a clear frequency shift ${\Delta}{\nu}_{0}$ between them.
With the opposite direction of $H$, the sign of ${\Delta}{\nu}_{0}$ is reversed (Fig. 2(c)).

In the following, we discuss the origin of the observed frequency shift ${\Delta}{\nu}_{0}$ between the spin waves propagating with the wave vector $+k$ and $-k$.
The effective Hamiltonian for the ferromagnets with chiral cubic crystal lattice can be described as $\mathscr{H}=\int d{\vec r} (E+H_{{\rm D}})$ with energy density $E$ given \cite{Kataoka} by
\begin{equation}
E= \frac{J}{2}({\nabla}{\vec S})^{2}-D{\vec S}{\cdot}[{\nabla}{\times}{\vec S}]-\frac{K}{2}\sum_{\substack{i}} S_{i}^{4} - \frac{\gamma \hbar}{V_{0}}{\mu}_{0}{\vec H}{\cdot}{\vec S},
\end{equation}
where $J$, $D$, and $K$ describe the magnitude of ferromagnetic exchange, DMI, and cubic anisotropy term, respectively.
$H_{\rm D}$ represents the magnetic dipole-dipole interaction, and ${\vec S}$, ${\gamma}$, $h=2{\pi}{\hbar}$, ${\mu}_{0}$, and $V_{0}$ are the dimensionless vector spin density, gyromagnetic ratio, Planck constant, magnetic permeability of vacuum, and the volume of formula unit cell, respectively.
In the case of the infinitely wide plate-shaped sample with the thickness $l$ and $H||k||[110]$ lying along the in-plane direction, the corresponding spin wave dispersion is deduced \cite{MagnetostaticWave,Kataoka} as
\begin{equation}
{\nu}=[{\rm sgn}({\vec k}{\cdot}{\vec H})] \frac{2DSV_{0}|k|}{h}+\frac{C_{\rm even}}{h},
\end{equation}
where $C_{\rm even}={\sqrt{(JSV_{0}k^{2}+{\tilde \Delta})(JSV_{0}k^{2}+{\tilde \Delta}+{\tilde K})}}$ with ${\tilde \Delta}=KV_{0}S^{3}+{\gamma}{\hbar}{\mu}_{0}H$ and ${\tilde K}={\gamma}{\hbar}{\mu}_{0}S[\frac{1-e^{-|k|l}}{|k|l}]-3KV_{0}S^{2}$.
Figures 1(j) and (k) exhibit the spin wave dispersion calculated by Eq. (2), with the parameters deduced for FeGe (through the detailed analysis of spin wave spectra as described below).
The dispersion is parabolic with its minimum at $|k| = D/J$, except for the $k \rightarrow 0$ region affected by the magnetic dipole-dipole interaction. 
The first and second terms in the right side of Eq. (2) are odd and even functions of $k$, respectively, indicating that only the former $k$-linear term originating from DMI can cause the asymmetry in spin wave dispersion.
Here, the magnitude of frequency shift ${\Delta} {\nu}_0 = {\nu}(+|k|) - {\nu}(-|k|)$ is described as
\begin{equation}
|{\Delta}{\nu}_{0}|=\frac{4DSV_{0}|k|}{h}.
\end{equation}
To experimentally confirm the predicted $k$-linear nature of $|{\Delta}{\nu}_{0}|$, we performed another measurement of $|{\Delta}{\nu}_{0}|$ using a similar waveguide pattern but with a different periodicity ${\lambda}$.
Figure 2(i) indicates the spectra of Im[${\Delta}L_{21}$] and Im[${\Delta}L_{12}$] measured with ${\lambda} = 24$ ${\mu}$m waveguides, while the previous data in Figs. 2(a) and (c) are measured with ${\lambda} = 12$ ${\mu}$m waveguides.
The obtained $|{\Delta}{\nu}_{0}|$ values are plotted against the central wavenumber $k_{\rm p} = 2{\pi}/{\lambda}$ of excited spin wave in Fig. 2(j), which shows that $|{\Delta}{\nu}_{0}|$ linearly increases with respect to $k_p$.
This confirms the validity of Eq. (3), i.e. the DMI origin of the observed frequency shift.

Importantly, Eq. (2) indicates that the sign of $\Delta {\nu}_{0}$ is also dependent on the sign of $D$ and $H$.
For FeGe, it has been reported that the left-handed (right-handed) chirality of crystals always host clockwise (counter-clockwise) helicity of skyrmion spin texture, for which the relevance of the sign difference of DMI has been discussed \cite{FeGe_LTEM,FeGe_helicity,Shibata}.
Figures 2(g) and (h) are the over-focused LTEM (Lorentz transmission electron microscopy) images obtained in the skyrmion-lattice state for left-handed and right-handed crystal pieces of FeGe, where the clockwise and counter-clockwise helicity of skyrmions appear as dark and bright spots reflecting the sign of Lorentz force acting on electron beam (Figs. 2(e) and (f)), respectively \cite{Shibata}.
We have performed the measurements of Im[${\Delta} L_{12}$] and Im[${\Delta} L_{21}$] for these left-handed (Figs. 2(a) and (c)) and right-handed (Figs. 2(b) and (d)) crystal pieces in the ferromagnetic state, and found that the sign of $\Delta \nu_{0}$ (and the associated sign of $D$) is clearly reversed between them.
These findings firmly establish the predicted coupling between the sign of DMI and skyrmion helicity.
As the origin of the presently observed $\Delta {\nu}_{0}$, we can eliminate the effects of the magnetic dipole-dipole interaction and the asymmetry of the surface on the basis of the above findings that $|\Delta {\nu}_{0}|$ satisfies the $k$-linear relation for two distinct wavenumbers of spin waves and that the opposite chirality of crystal reverses the sign of $\Delta {\nu}_{0}$.

On the basis of Eq. (3), the magnitude of $D$ can be directly evaluated from the observed $|{\Delta}{\nu}_{0}|$, using the $S$ value deduced from the saturated magnetization $M_{\rm S}= {\gamma}{\hbar}S/V_{0}$ in the $M$-$H$ profile (not shown).
The other two magnetic parameters $J$ and $K$ can be further determined so as to reproduce the $H$ dependence of magnetic resonance frequency (Fig. 1(d)) in the ferromagnetic state using Eq. (2), and the critical magnetic field $H_{c}$ given \cite{MagnetostaticWave,Kataoka} by
\begin{equation}
\frac{\gamma \hbar}{V_{0}}{\mu}_{0}H_{c} = \frac{D^{2}S}{J} + \frac{KS^{3}}{2}-\frac{9JK^{2}S^{5}}{16D^{2}}.
\end{equation}
The red solid line in Fig. 1(d) is the fitting curve based on Eq. (2), and the magnetic parameters for FeGe determined from the spin wave spectra are listed in Table 1.

Likewise, we have investigated the propagation character of spin wave for Co$_{8}$Zn$_{8}$Mn$_{4}$ and Co$_{9}$Zn$_{9}$Mn$_{2}$.
Figures 3(a) and (c) indicate the Im[${\Delta}L_{21}$] and Im[${\Delta}L_{12}$] spectra measured for these compounds at 200 K under $H = +140$ mT, i.e. in the collinear ferromagnetic spin state.
For the both compounds, the spin wave propagating with the wave vector $+k$ and $-k$ shows a clear frequency shift ${\Delta}{\nu}_{0}$, whose sign is confirmed to be reversed for opposite direction of $H$ (Figs. 3(b) and (d)).
The observed $|{\Delta}{\nu}_{0}|$ value allows us to directly estimate the magnitude of DMI following Eq. (3), and we can determine all the relevant magnetic parameters so as to reproduce the experimentally observed $H_{c}$ value and $H$-dependence of magnetic resonance frequency in the ferromagnetic state (Figs. 1(e) and (f)) as in case of FeGe (See Supplemental Material \cite{SM} for details).
The obtained magnetic parameters for Co$_{8}$Zn$_{8}$Mn$_{4}$ and Co$_{9}$Zn$_{9}$Mn$_{2}$ are summarized in Table 1.

On the basis of these magnetic parameters determined from the spin wave spectra, we can elucidate the microscopic origin of helimagnetism and skyrmion formation in FeGe, Co$_{8}$Zn$_{8}$Mn$_{4}$, and Co$_{9}$Zn$_{9}$Mn$_{2}$.
When the DMI dominantly contributes to the emergence of helimagnetism, the magnetic modulation period ${\lambda}_{\rm h}$ in the helical or skyrmion-lattice spin state should be given \cite{Kataoka} as
\begin{equation}
{\lambda}_{\rm h}=2{\pi}J/D.
\end{equation}
To testify the validity of this model for each compound, the values of $2{\pi}J/D$ deduced from the present PSWS measurements (Table 1) are plotted against the actual ${\lambda}_{\rm h}$ values reported previously by the small angle neutron scattering (SANS) experiments \cite{FeGe_SANS,Tokunaga} in Fig. 4.
For all compounds, the values of $2{\pi}J/D$ and ${\lambda}_{\rm h}$ show a good agreement with each other, which confirms the validity of Eq. (5).
The above results quantitatively prove that the helimagnetism and the associated skyrmion formation originate dominantly from the interplay between DMI and ferromagnetic exchange interaction, rather than other potential mechanisms such as magnetocrystalline anisotropy and dipole-dipole interaction, in these room-temperature chiral-lattice magnets.
The good reproduction of $\lambda_{\rm h}$ also implies the accuracy of $D$ value estimated from the frequency shift $|\Delta \nu_0|$ in the PSWS measurements.

Note that such DMI-induced frequency shift between counter-propagating spin waves is observable even at 300 K for Co$_9$Zn$_9$Mn$_2$ (Fig. 3(c), inset), which demonstrates the first clear observation of this phenomenon at room temperature in bulk metallic single-phase compounds.
This can be viewed as a function of spin-wave diode owing to the chiral crystal structure, and may serve as an unique building element for the spintronics based on the concept of spin-wave spin current \cite{Kajiwara}.

In summary, we experimentally identified all the relevant magnetic parameters for room-temperature chiral-lattice magnets FeGe and Co-Zn-Mn alloys, by investigating the propagation character of spin wave.
On the basis of the observed frequency shift between counter-propagating spin waves, the magnitude and sign of DM interaction were directly evaluated. 
Combined with the real-space observation of spin texture, the sign of DM interaction is confirmed to be coupled with the skyrmion helicity.
The magnetic parameters obtained from spin wave spectra quantitatively account for the reported skyrmion size and its material variation, which proved that the DM interaction plays a decisive role in the helimagnetism and skyrmion formation in these compounds.
The propagating spin-wave spectroscopy can thus be an efficient tool to study DM interaction in bulk single-phase compounds, and the present results will provide a fundamental basis for the further parameter tuning and material search to obtain the desirable size and stability of room-temperature skyrmions toward their potential storage application.

The authors thank K. Kondou, Y. Otani, T. Koretsune, F. Kagawa, N. Ogawa, Y. Okamura, and A. Kikkawa for enlightening discussions and experimental help.
This work was partly supported by the Mitsubishi Foundation and Grants-In-Aid for Scientific Research (Grants No. 15H05458, No. 16K13842, and No. 17H05186) from JSPS.

\begin{figure}
\begin{center}
\includegraphics[width=9.5cm,keepaspectratio]{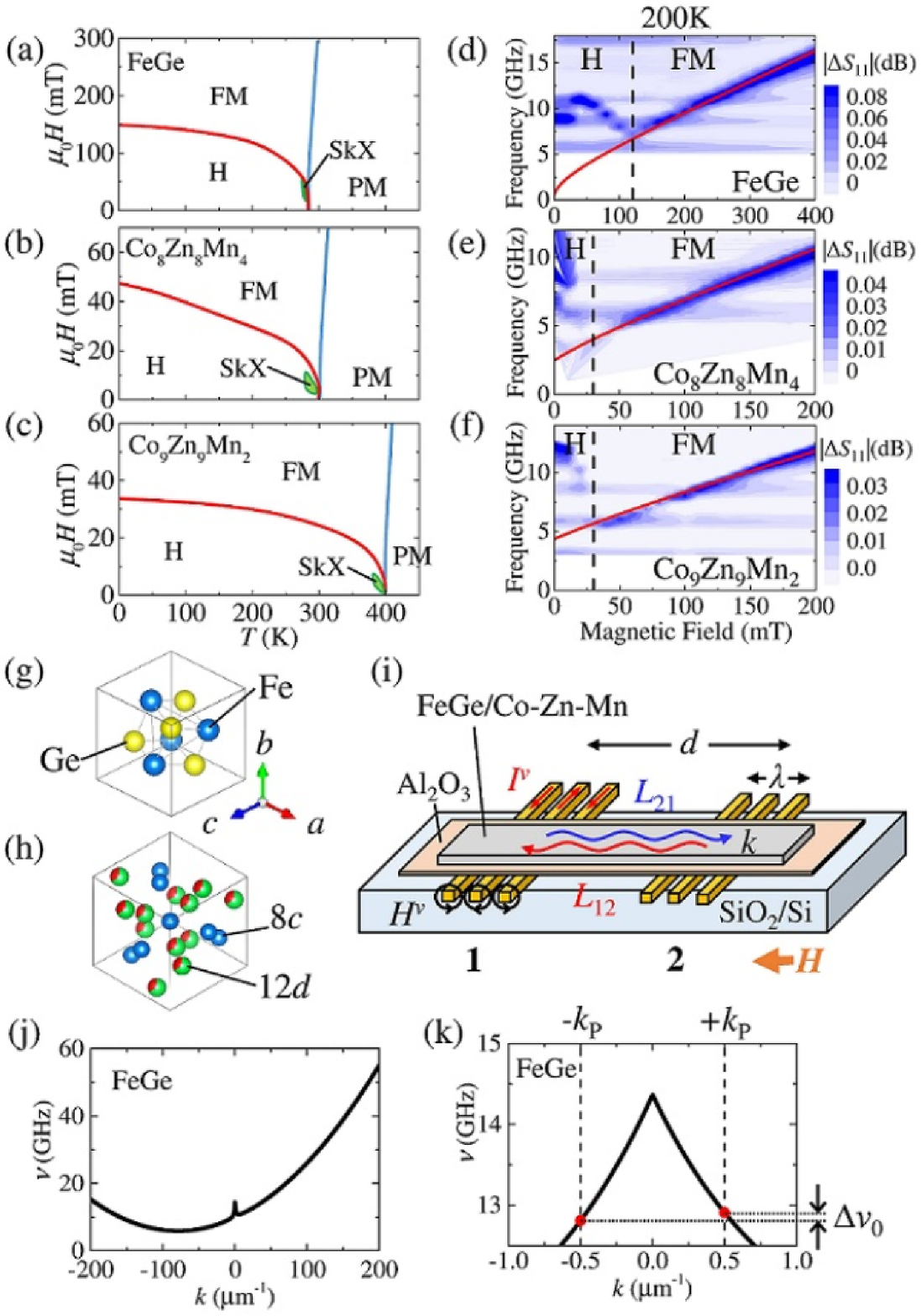}
\end{center}
\caption{(Color online) (a)-(c) $H$-$T$ phase diagrams previously reported in Refs.\cite{acchi_FeGe,Tokunaga} and (d)-(f) magnetic field dependence of microwave absorption spectra $|{\Delta}S_{11}|$ measured at 200 K, for FeGe, Co$_{8}$Zn$_{8}$Mn$_{4}$ and Co$_{9}$Zn$_{9}$Mn$_{2}$. 
H, SkX, FM, and PM correspond to helical, skyrmion lattice, ferromagnetic, and paramagnetic spin states, respectively.
Black dashed lines, red solid lines, and background color in (d)-(f) represent the critical magnetic field $H_{c}$, theoretical fitting based on Eq. (2), and magnitude of absorption strength $|{\Delta} S_{11}|$, respectively.
(g) and (h) Crystal structures of FeGe and Co-Zn-Mn alloys. The latter contains two kinds of crystallographic sites, $8c$ and $12d$, which are mainly occupied by Co and Zn/Mn atoms, respectively, but with some randomness \cite{CoZnMn}.
(i) Schematic illustration of the device structure for spin-wave spectroscopy experiments. 
(j) Spin wave dispersion in the collinear ferromagnetic state calculated for FeGe based on Eq. (2) with the material parameters in Table 1 as well as ${\gamma}/2{\pi}= 28$ GHz T$^{-1}$, ${\mu}_{0}M_{\rm S} = 0.14$ T, ${\mu}_{0}H = 0.35$ T, and $V_{0} = 104$ $\rm{\AA}^{3}$.
(k) Magnified view of the dispersion around $k=0$.
}
\label{Fig1}
\end{figure}

\begin{figure}
\begin{center}
\includegraphics[width=9.5cm,keepaspectratio]{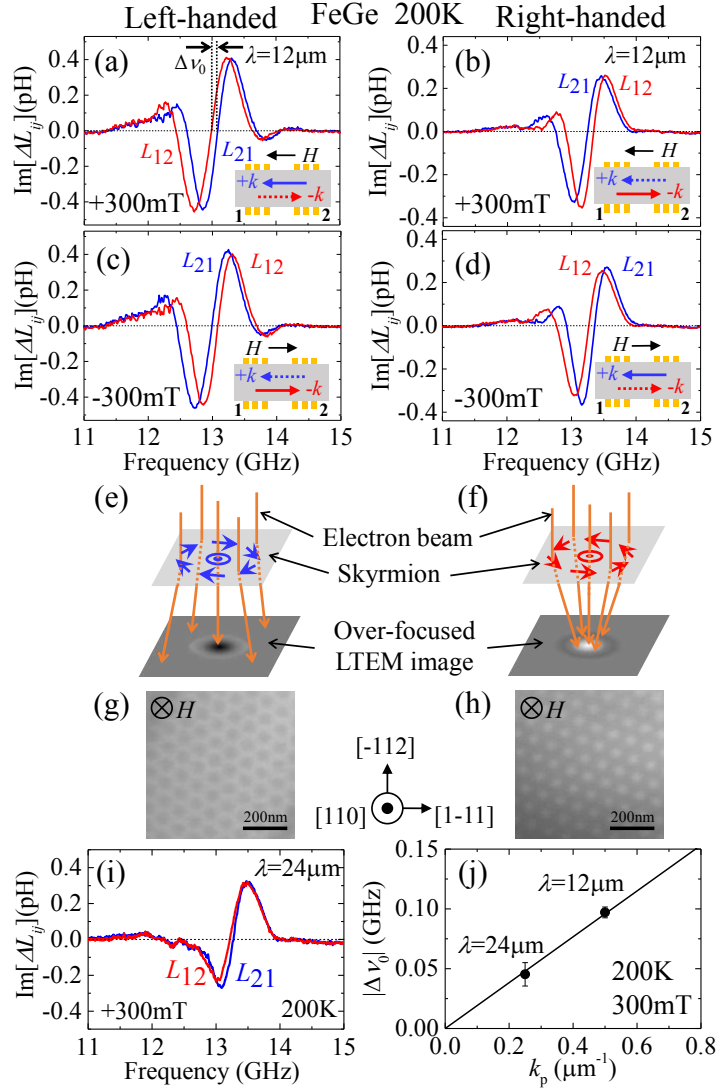}
\end{center}
\caption{(Color online) Nonreciprocal spin-wave propagation in FeGe.
(a)-(d) Imaginary part of mutual inductance ${\Delta}L_{21}$ and ${\Delta}L_{12}$ measured with various combination of magnetic field direction ($H ={\pm}300$ mT) and crystallographic chirality (left-handed or right-handed).
Here, $\Delta L_{21}$ and $\Delta L_{12}$ represent the spin wave propagating with wave vector $+k$ and $-k$.
The definition of frequency shift ${\Delta}{\nu}_{0}$ is indicated in (a).
(e) and (f) Schematic illustration of the relationship between the skyrmion helicity and the resultant contrast in the over-focused LTEM image.
(g) and (h) Over-focused LTEM images of skyrmions, obtained for left-handed and right-handed crystal pieces of FeGe, respectively \cite{LTEMcondition}.
(i) Imaginary part of ${\Delta}L_{21}$ and ${\Delta}L_{12}$ measured with the waveguide pattern of ${\lambda} = 24$ ${\mu}$m, whose $\lambda$ value is doubled compared to the one ($\lambda = 12$ $\mu$m) used in (a)-(d).
(j) The magnitude of frequency shift ${\Delta}{\nu}_{0}$ as a function of $k_{\rm p} = 2{\pi} / {\lambda}$.
A solid line represents the liner fitting to the data.}
\label{Fig2}
\end{figure}

\begin{figure}
\begin{center}
\includegraphics[width=11.5cm,keepaspectratio]{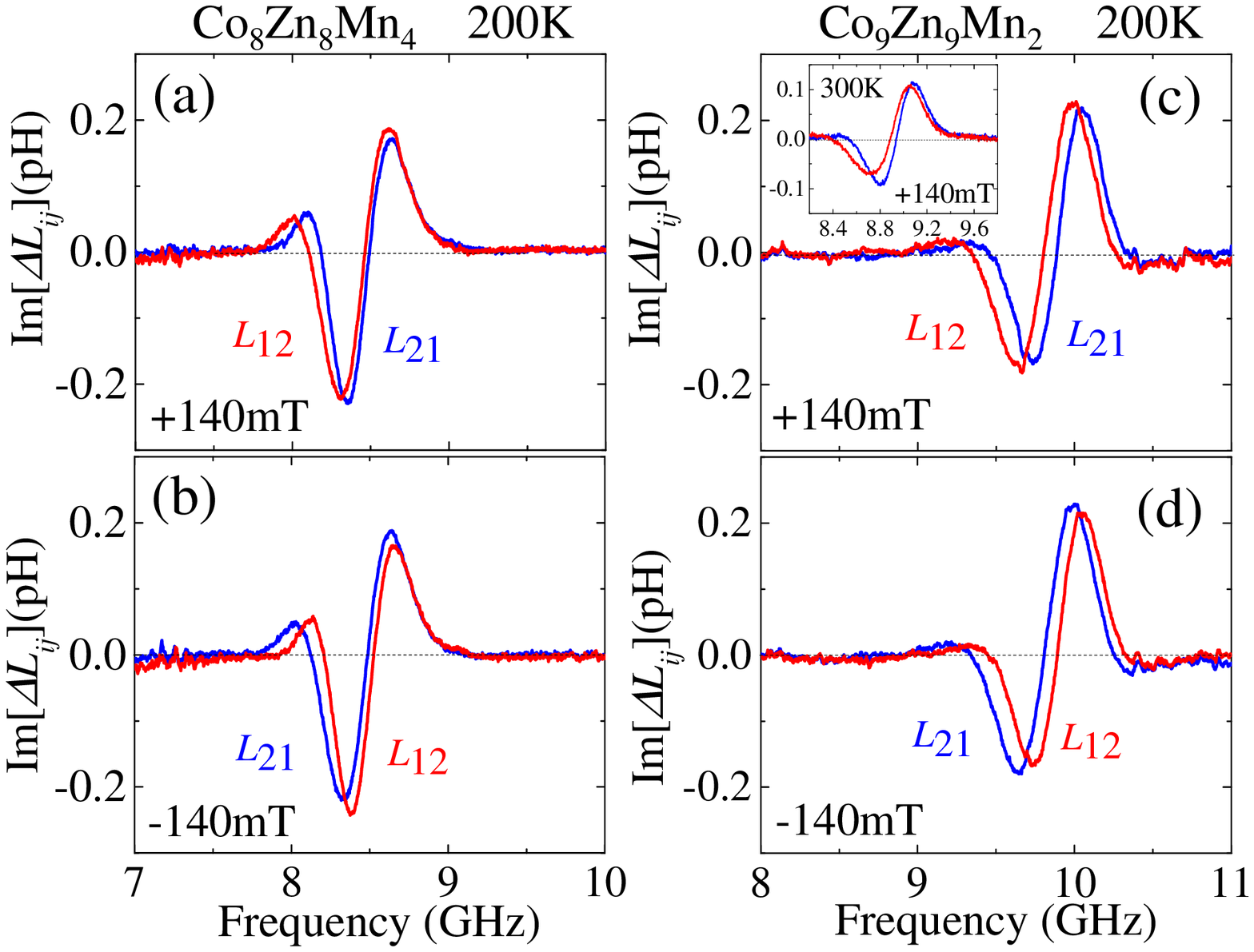}
\end{center}
\caption{(Color online) Nonreciprocal spin-wave propagation in Co-Zn-Mn alloys.
Imaginary part of mutual inductance ${\Delta}L_{21}$ and ${\Delta}L_{12}$ at 200 K with $H = \pm 140$ mT, measured for (a),(b) Co$_8$Zn$_8$Mn$_4$ and (c),(d) Co$_{9}$Zn$_{9}$Mn$_{2}$.
The inset in (c) represents the corresponding data measured at 300 K.}
\label{Fig3}
\end{figure}

\begin{figure}
\begin{center}
\includegraphics[width=7.0cm,keepaspectratio]{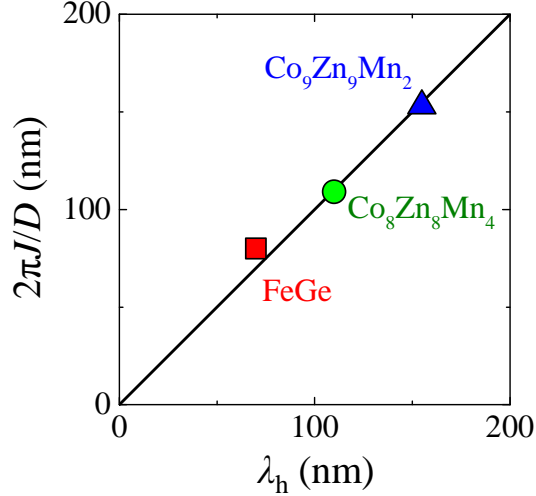}
\end{center}
\caption{(Color online) Comparison between the $2{\pi}J/D$ value determined from the present PSWS measurements (Table 1) and the actual helical spin modulation period ${\lambda}_{\rm h}$ determined previously by SANS experiments \cite{FeGe_SANS,Tokunaga}. 
Solid line represents the relationship expected from Eq. (5) in the text.}
\label{Fig4}
\end{figure}

\newpage
\begin{table}
\begin{center}
\caption{Summary of magnetic parameters for FeGe and Co-Zn-Mn alloys, obtained through the present PSWS experiment at 200 K.}
\begin{tabular}{|c|c|c|c|c|c|c|c|} \hline
 &$D$ (J/m$^{2}$) &$J$ (J/m) &$K$ (J/m$^{3}$) &$S$ \\ \hline
FeGe & $2.8{\times}10^{-3}$ &$3.6{\times}10^{-11}$ &$1.0{\times}10^{4}$ &0.45 \\ \hline
Co$_{8}$Zn$_{8}$Mn$_{4}$ & $5.3{\times}10^{-4}$ &$9.2{\times}10^{-12}$ &$4.0{\times}10^{4}$ &0.58 \\ \hline
Co$_{9}$Zn$_{9}$Mn$_{2}$ & $1.2{\times}10^{-3}$ &$2.8{\times}10^{-11}$ &$4.0{\times}10^{4}$ &0.75 \\ \hline
\end{tabular}
\end{center}
\end{table}


\begin{thebibliography}{999}
\bibitem{LTEMcondition}The LTEM observations were performed at 260 K under a magnetic field of 100 mT along the [110] axis \cite{SM}.
\end{thebibliography}
\end{document}


\title{Supplemental Material: \\
Spin-wave spectroscopy on Dzyaloshinskii-Moriya interaction\\
in room-temperature chiral magnets hosting skyrmions}
\author{R. Takagi$^{1}$, D. Morikawa$^{1}$, K. Karube$^{1}$, N. Kanazawa$^{2}$, K. Shibata$^{1}$,\\
G. Tatara$^{1}$, Y. Tokunaga$^{3}$, T. Arima$^{1,3}$, Y. Taguchi$^{1}$, Y. Tokura$^{1,2}$, and S. Seki$^{1}$}
\affiliation{$^{1}$RIKEN Center for Emergent Matter Science(CEMS), Wako 351-0198, Japan,\\
$^{2}$Department of Applied Physics, University of Tokyo, Bunkyo-ku 113-8656, Japan, $^{3}$Department of Advanced Materials Science, University of Tokyo, Kashiwa 277-8561, Japan}
\date{\today}

\maketitle

\section*{Experimental details}
Our measurement and data analysis procedures associated with propagating spin-wave spectroscopy follow the method described in Refs. \cite{Seki,Vlaminck1,Vlaminck2}.
To investigate the propagation character of spin wave, we fabricated the device structures as summarized in Figs. S1(a) and (b).
A pair of coplanar waveguides consisting of Au(195 nm)/ Ti(5 nm) were patterned on the oxidized silicon substrates by photolithography, and 100-nm-thick of insulating Al$_{2}$O$_{3}$ was deposited on top of them.
Then, a plate-shaped single crystal of FeGe or Co-Zn-Mn alloys with typical size from 70 ${\mu}$m $\times$ 15 ${\mu}$m $\times$ 1 ${\mu}$m to 60 ${\mu}$m $\times$ 10 ${\mu}$m $\times$ 1 ${\mu}$m was extracted from the original bulk crystal pieces and placed across the waveguides with W deposition at both edges of the crystal by focused ion beam (FIB) microfabrication technique.
Each waveguide was connected with a vector network analyzer (VNA) through a coaxial cable and ground-signal-ground (GSG) microprobe.
An oscillating electric current $I^{\nu}$ of gigahertz frequency ${\nu}$ was injected into one of the waveguides via VNA.
This $I^{\nu}$ generated oscillating magnetic field $H^{\nu}$ and excited spin waves in the sample.
The propagating spin wave generated an additional magnetic flux on the waveguides, which induced an oscillating electric voltage $V^{\nu}$ following Faraday's law.
The complex spectrum of scattering parameters ($S_{nm}$($\nu$, $H$) with $n$ and $m$ being the port number used for the excitation and detection of spin wave, respectively) was measured by VNA at various magnitudes of external magnetic field, and converted into the impedance spectrum $Z_{nm}({\nu},H)$ defined as $V_{n}^{\nu} = \sum_{m} Z_{nm}({\nu}) I_{m}({\nu})$ assuming the characteristic impedance $Z_0 = 50$ ${\Omega}$ \cite{FieldsAndWaves}.
The spectrum at a reference field $H$ = $H_{\rm ref}$, for which no magnetic resonant behavior occurs within the frequency range of interest, was used as the background.
By subtracting the referential impedance from the resonant one, the spin-wave contribution to the inductance spectrum was deduced as ${\Delta}L_{nm}(\nu, H) = [Z_{nm}(\nu, H)-Z_{nm}(\nu,H_{\rm ref})]/(i2{\pi}{\nu})$.
For the analysis of the propagation character of spin wave, we mainly investigated the mutual inductance $\Delta L_{12}$ and $\Delta L_{21}$ between two waveguides.
To evaluate the local excitation strength of spin wave, we employed $\Delta S_{11} (\nu, H) = S_{11} (\nu, H) - S_{11} (\nu, H_{\rm ref})$.
In the 	present work, the wave vector ${\vec k}$ of spin wave was along the $[110]$ axis for FeGe and along the $[100]$ axis for Co-Zn-Mn alloys, respectively.
External magnetic field was always applied parallel to $k$ (i.e. $H||k$), which corresponds to the geometry of magnetostatic backward volume waves \cite{MagnetostaticWave}.

Figure S1(c) is a side view of the coplanar waveguide pattern employed in the present study.
Each waveguide consists of one signal line and two ground lines, terminated with a short circuit.
When it is connected to the VNA through the GSG microprobe, the input current densities for the signal and ground line are $I^{\nu}$ and $-I^{\nu}/2$, respectively.
The distribution of the wave number $k$ of the excited spin wave is given by the Fourier transform of the waveguide pattern \cite{Vlaminck1,Vlaminck2} as plotted in Fig. S1(d).
A main peak of wave number for the waveguide pattern with the periodicity ${\lambda} = 12$ ${\mu}$m and ${\lambda} = 24$ ${\mu}$m is at $k_{\rm p} = 0.50$ ${\mu}$m$^{-1}$ and $k_{\rm p} = 0.25$ ${\mu}$m$^{-1}$, respectively.
These satisfy the relationship $k_{\rm p}$ ${\sim}$ $2{\pi}/{\lambda}$.
While the higher order peaks are found in the larger $k$ regions, the amplitude of the second largest one is $1/7$ times as large as the main peak.
Therefore, we analyzed the ${\Delta}L_{nm}$ spectra assuming only the main peak contribution centered at $k_{\rm p}$.

The LTEM (Lorentz transmission electron microscopy) study was carried out using a conventional TEM (JEM-2100F) at an acceleration voltage of 200 kV.
By using the FIB microfablication technique, thin-plate samples with typical thickness of 150 nm were prepared from left-handed and right-handed crystal pieces of FeGe, which are the same ones as employed for spin wave measurements.
Their crystal orientation was confirmed using electron diffraction patterns.
In LTEM observations, the magnetic structures can be imaged as convergences (bright contrast) or divergences (dark contrast) of the electron beam on the defocused (under- or over-focused) image planes.
Sample temperature was kept at 260 K, controlled by a liquid-nitrogen specimen holder, and a magnetic field of 100 mT was applied perpendicular to the specimen plate.
According to Ref. \cite{FeGe_LTEM}, the phase diagram of thin-plate sample is dependent on the sample thickness, and the skyrmion phase is generally stabilized for a wider temperature range in a thinner sample.
Therefore, the $H$-$T$ phase diagram of the thin-plate sample for LTEM observations should be slightly different from that of the bulk sample shown in Fig. 1(a) of the main text.

\section*{Spin wave dispersion}
The effective Hamiltonian for the ferromagnets with chiral cubic symmetry can be described as $\mathscr{H}=\int d{\vec r} (E+H_{{\rm D}})$ with energy density $E$ \cite{Kataoka}
\begin{equation}
E= \frac{J}{2}({\nabla}{\vec S})^{2} - D{\vec S}{\cdot}[{\nabla}{\times}{\vec S}]-\frac{K}{2}\sum_{\substack{i}} S_{i}^{4} - \frac{\gamma \hbar}{V_{0}}{\mu}_{0}{\vec H}{\cdot}{\vec S},
\end{equation}
where $J$, $D$, and $K$ describe the magnitude of ferromagnetic exchange, DMI, and cubic anisotropy term, respectively.
$H_{\rm D}$ represents the magnetic dipole-dipole interaction, and ${\vec S}$, ${\gamma}$, $h=2{\pi}{\hbar}$, ${\mu}_{0}$, and $V_{0}$ are the dimensionless vector spin density, gyromagnetic ratio, Planck constant, magnetic permeability of vacuum, and the volume of formula unit cell, respectively.
When the infinitely wide plate-shaped sample with the thickness $l$ is assumed and $H||k$ lies along the in-plane direction, the corresponding spin wave dispersion is given \cite{MagnetostaticWave,Kataoka} as
\begin{equation}
{\nu}=[{\rm sgn}({\vec k}{\cdot}{\vec H})] \frac{2DSV_{0}|k|}{h}+\frac{C_{\rm even}}{h},
\end{equation}
with $C_{\rm even}={\sqrt{(JSV_{0}k^{2}+{\tilde \Delta})(JSV_{0}k^{2}+{\tilde \Delta}+{\tilde K})}}$.
In case of $H{\parallel}k{\parallel}[110]$, ${\tilde \Delta}_{[110]}=KV_{0}S^{3}+{\gamma}{\hbar}{\mu}_{0}H$ and ${\tilde K}_{[110]}={\gamma}{\hbar}{\mu}_{0}S[\frac{1-e^{-|k|l}}{|k|l}]-3KS^{2}$.
Likewise, in the configuration $H||k||[100]$, ${\tilde \Delta}_{[100]}=2KV_{0}S^{3}+{\gamma}{\hbar}{\mu}_{0}H$ and ${\tilde K}_{[100]}={\gamma}{\hbar}{\mu}_{0}S[\frac{1-e^{-|k|l}}{|k|l}]$.
For both configurations, $C_{\rm even}$ is a even function against $k$, and therefore only the first $k$-linear term in the right side of Eq. (S2) can cause an asymmetry in spin wave dispersion.

Based on Eq. (S1), the critical magnetic field ${\mu}_{0}H_{c}$ (required to induce transition from helical to ferromagnetic spin state) is deduced \cite{Kataoka} as
\begin{equation}
\frac{\gamma \hbar}{V_{0}}{\mu}_{0}H_{c} = \frac{D^{2}S}{J} + {\tilde H},
\end{equation}
where ${\tilde H}=\frac{KS^{3}}{2}-\frac{9JK^{2}S^{5}}{16D^{2}}$ for $H||[110]$ and ${\tilde H}=-2KS^{3}$ for $H||[100]$.
To analyze the obtained spin wave spectra, we used the above relationships for corresponding measurement geometries.

\section*{Magnetic field dependence of frequency shift}
To confirm the reliability of the $D$-value estimated from $|\Delta \nu_{0}|$, we investigated the magnetic field dependence of the magnitude of frequency shift $|\Delta \nu_{0}|$ between the spin waves propagating with the wave vector $+k$ and $-k$ (Fig. S2).
In the collinear ferromagnetic state above $H_{c}$, $|\Delta \nu_{0}|$ shows a nonzero value and remains almost $H$-independent.
In the helimagnetic state below $H_{c}$, on the other hand, the value of $|\Delta \nu_{0}|$ vanishes, which is in accord with the theoretical prediction and probably reflects the folding back of magnetic Brillouin zone with the helical modulation period \cite{Kataoka}.
For the estimation of $D$-value in the main text, we employed the average $|\Delta \nu_{0}|$ value for $H/H_{c}>2$ to avoid the slight distortion of spectrum shape around the magnetic phase boundary.
The standard deviation of $|\Delta \nu_{0}|$ is plotted as the error bars on Fig. 2(j) in the main text.

\newpage
\begin{figure}
\begin{center}
\includegraphics[width=8cm,keepaspectratio]{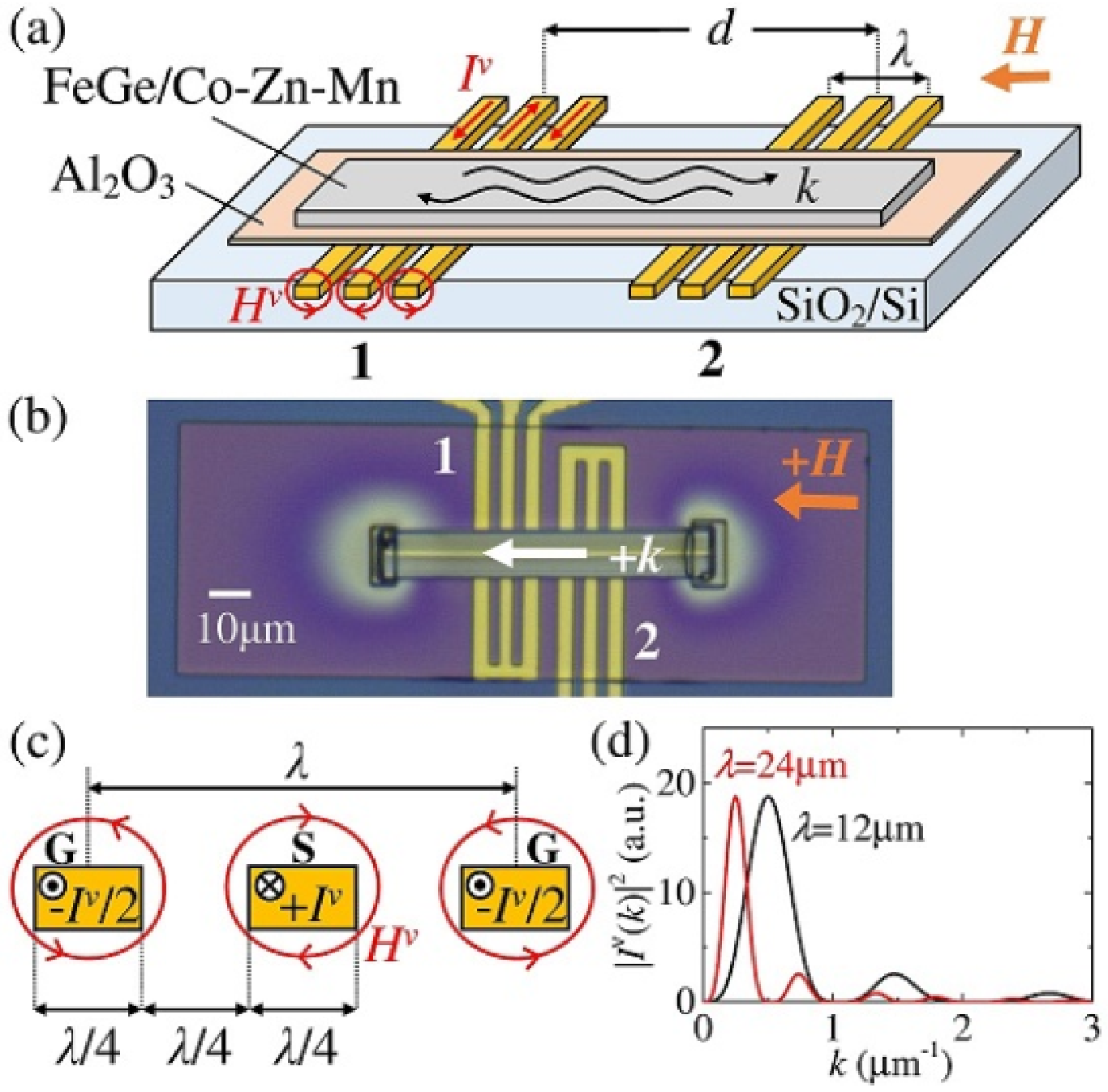}
\end{center}
\caption{(Color online) (a) Schematic illustration and (b) the optical microscope image of the device structure for spin-wave spectroscopy experiments. 
(c) Side view of the coplanar waveguide consisting of one signal line (S) and two ground lines (G) along with the associated current density distribution.
(d) Wave number distribution of excitation current $I^{\nu}(k)$, obtained by the Fourier transform of the waveguide pattern with ${\lambda}=12$ ${\mu}$m (black) and ${\lambda}=24$ ${\mu}$m (red).
}
\end{figure}

\begin{figure}
\begin{center}
\includegraphics[width=7cm,keepaspectratio]{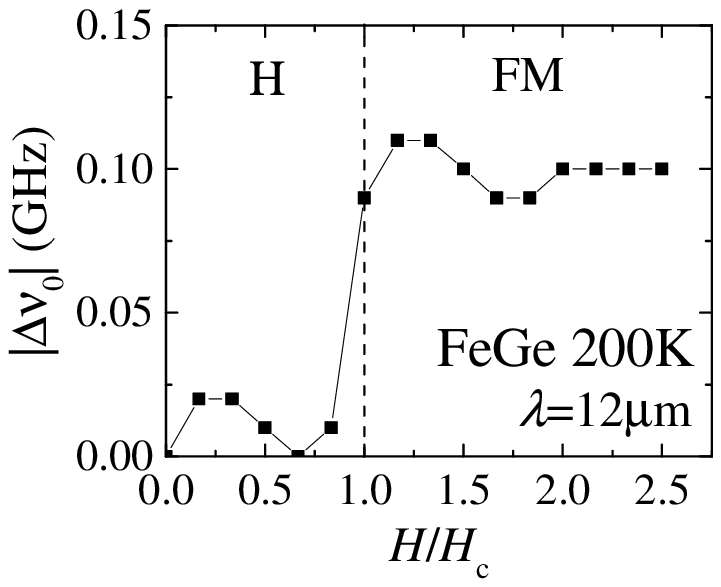}
\end{center}
\caption{(Color online) Magnetic field dependence of the magnitude of frequency shift $|\Delta \nu_{0}|$ measured for FeGe with the waveguide patterns of ${\lambda}=12$ $\mu$m.
H and FM correspond to helical and ferromagnetic spin states, respectively.
}
\end{figure}